\documentstyle[preprint,aps,epsfig]{revtex}

\tightenlines
\def \beq{\begin{equation}}  \def \eeq{\end{equation}}
\def \beqa{\begin{eqnarray}} \def\eeqa{\end{eqnarray}}
\def \nonb{\nonumber}
\begin{document}
\draft
\preprint{AS-ITP-99-14}
\title{\Large\bf $F^0-\bar{F}^0 $ Mixing and CP Violation \\
in the General Two Higgs Doublet Model}
\author{Y.L Wu and Y.F Zhou}
 \address{\small Institute of Theoretical Physics, Chinese Academy of Sciences, \\ 
Beijing 100080,  China}
\maketitle

\begin{abstract}
     A phenomenological analysis of the general two Higgs doublet model
   is presented. Possible constraints of the Yukawa couplings are obtained from
   the $K^0-\bar{K}^0 $, $B^0-\bar{B}^0 $ and $D^0-\bar{D}^0 $ mixings. A much larger 
   $D^0-\bar{D}^0 $ mixing than the standard model prediction is possible.
   It is shown that the emerging of various new sources 
   of CP violation in the model could strongly affect the determination of the
   unitarity triangle. It can be helpful to look for a signal of new physics by 
   comparing the extracted angle $\beta$ from two different ways, such as from 
  the process $B\rightarrow J/\psi K_{S}$ and from fitting the quantities $|V_{ub}|$, 
$\Delta m_{B}$ and $\epsilon$.   
\end{abstract}

\section{Introduction}

In the standard model(SM) of an electroweak $SU(2)_L \times U(1)_Y$ gauge theory
with only one Higgs doublet, the only source of CP violation comes from
the complex Yukawa coupling between Higgs and fermion fields\cite{KM}.
Since the Higgs sector of SM is not well understood yet,
many possible extensions of SM have been proposed\cite{mode12}. 
One of the simplest extensions of the SM
is to simply add one Higgs doublet without imposing the {\it ad hoc} discrete symmetries. 
For the convenience of mention in our following discussions, 
we may call such a minimal extension of the standard model with generally adding an extra Higgs doublet 
 as an S2HDM\cite{TH1,TH2,TH3,TH4,WW,YLW,soni,BCK,KSW} and 
assume CP violation solely originating from the Higgs potential\cite{TDL,WW,YLW}.
The most general Yukawa coupling and Higgs potential can be written as:
\beqa
L_Y=\bar{Q}_L (\Gamma^U_1 \tilde{\phi}_1 +\Gamma^U_2 \tilde{\phi}_2)U_R 
    +\bar{Q}_L (\Gamma^D_1 \phi_1 +\Gamma^D_2 \phi_2)D_R
\eeqa
and:
\beqa
V(\phi_1,\phi_2)=&-&\mu_1^2 \phi_1^{\dagger} \phi_1
                 -\mu_2^2 \phi_2^{\dagger} \phi_2
                 -(\mu_{12}^2  \phi_1^{\dagger} \phi_2 +h.c) \nonumber \\
                 &+&\lambda_1 (\phi_1^{\dagger} \phi_1)^2
                 +\lambda_2 (\phi_2^{\dagger} \phi_2)^2  
                 +\lambda_3 (\phi_1^{\dagger} \phi_1 \phi_2^{\dagger} \phi_2)
                 +\lambda_4 (\phi_1^{\dagger} \phi_2 \phi_2^{\dagger} \phi_1) \nonb\\
                 &+&\frac{1}{2}[\lambda_5 (\phi_1^{\dagger} \phi_2)^2 +h.c]
                 +[(\lambda_6 \phi_1^{\dagger} \phi_1
                    +\lambda_7 \phi_2^{\dagger} \phi_2)
                    (\phi_1^{\dagger} \phi_2) +h.c]
\eeqa
 The major issue with respect to the
two Higgs doublet model is that it allows flavor changing neutral current
(FCNC) at tree level, which must be strongly suppressed in $K^0-\bar{K}^0$
and $B^0-\bar{B}^0$ mixing processes. In order to prevent FCNC from tree level, 
an $ad \ hoc$ discrete symmetry is often imposed 
\beqa  
 \phi_1 \rightarrow -\phi_1 & \mbox{\ and \ } & \phi_2 \rightarrow \phi_2 \nonb \\
 U_{R_i} \rightarrow -U_{R_i} & \mbox{\ and \ }& D_{R_i} \rightarrow \mp D_{R_i}
\eeqa
Thus, one obtains the so called mode I and model II, which depends on whether the up-
type and down-type quarks are coupled to the same or different Higgs doublet respectively
\cite{mode12}. Once the discrete symmetry is adopted, the factor $\mu_{12},
\lambda_6$ and $\lambda_7$ in Eq.(2) mush vanish. As a result,  no CP violation
 can occur  from $V(\phi)$. Thus the only source of CP violation
is the complex Yukawa couplings, which lead to a phase in the Cabibbo-Kobayashi-
Maskawa(CKM) quark mixing matrix.
	
	In contrast, one can replace the discrete symmetry by an approximate global
family symmetry\cite{TH2,TH3,WW,YLW} , thus the suppression of FCNC can be explained    
via the smallness of the off-diagonal terms.
 Furthermore, when abondoning the discrete symmetries,  one can obtain, after spontaneous
 symmetry breaking, rich sources of CP violation from a single relative phase 
between the two vacuum 
 expectation values of Higgs fields. It has been shown\cite{WW,YLW} that even when the 
CKM matrix is real, the single phase arising from the spontaneous symmetry 
 breaking can provide enough CP violation to meet the experimental measurments. 
 One of particular important observations is a new source of CP
 violation in charged Higgs boson interactions, which is independent of
 the CKM phase and can lead to a value of $\epsilon'/ \epsilon$ 
  as large as $10^{-3}$\cite{TH3}.  In the S2HDM,
the two Higgs fields have, in general, the vacuum expectation values:
\beqa  <\phi_1^0>&=&\frac{v}{\sqrt{2}} \cos\beta e^{i\delta}  \nonb \\
      <\phi_2^0>&=&\frac{v}{\sqrt{2}} \sin\beta 
\eeqa
It is natural to use a suitable basis:
\beqa
H_1 &=&\cos\beta \phi_1 e^{-i\delta} + \sin\beta \phi_2 \nonb \\
H_2 &=&\sin\beta \phi_1 e^{-i\delta} - \cos\beta \phi_2 
\eeqa
such that:
\beqa
H_1 &=&\left( \begin{array}{c}
              0  \\ \frac{1}{\sqrt{2}}(v+ \rho)
            \end{array} 
    \right) \nonb \\
H_2 &=&\left(  \begin{array}{c}
              H^+  \\ \frac{1}{\sqrt{2}}(R+iI)  
             \end{array}
     \right),
\eeqa
where $H^0,R,I$ are real Higgs bosons. The three neutral scalars $\hat{H}_k^0 \equiv 
(R, \rho ,I)$ can be rotated into mass engenstates $h_k^0 \equiv (h,H^0,A)$ via  
orthogonal matrix $O^H$.
\beq  \hat{H}_k^0=O^H_{kl} H^0_l \eeq

   From approximate global family symmetries,  we know that the Yukawa coupling matrices $\Gamma^F_i$
in Eq.(1) have small off-diagonal elements, typically between 0.01 and 0.2 in order
 to meet the constraint  of FCNC from $K^0-\bar{K}^0, B^0-\bar{B}^0$ mixing.
The Yukawa interaction can be rewritten as\cite{WW}:
\beq
L_Y=(L_1+L_2)(\sqrt{2} G_F)^{1/2}
\eeq
with
\beqa
L_1=&\sqrt{2}&\left(
    H^+ \sum\limits^{3}_{i,j} \xi_{d_j} m_{d_j} V_{ij} \bar{u}_L^i d_R^j
   -H^- \sum\limits^{3}_{i,j} \xi_{u_j} m_{u_j} V_{ij}^{\dagger} \bar{d}_L^i u_R^j
   \right)  \nonb \\
   &+&H^0 \sum\limits^{3}_{i}(m_{u_i} \bar{u}_L^iu_R^i + m_{d_i} \bar{d}_L^i d_R^i) \nonb \\
   &+&(R+iI)\sum\limits_i^3 \xi_{d_i} m_{d_i} \bar{d_L^i} d_R^i
     +(R-iI)\sum\limits_i^3 \xi_{u_i} m_{u_i} \bar{u_L^i} u_R^i +h.c \\
L_2=&\sqrt{2}&\left(
    H^+ \sum\limits^{3}_{i,j'\not=j}  V_{ij'}\mu_{j'j}^d \bar{u}_L^i d_R^j
   -H^- \sum\limits^{3}_{i,j'\not=j}  V_{ij}^{\dagger}\mu_{j'j}^u \bar{d}_L^i u_R^j
   \right)  \nonb \\   
   &+&(R+iI)\sum\limits_{i\not=j}^3 \mu_{ij}^d  \bar{d_L^i} d_R^j
   +(R-iI)\sum\limits_{i\not=j}^3 \mu_{ij}^u  \bar{u}_L^i u_R^j+h.c,
\eeqa         
where $L_1$ has no flavor-changing effects other than that expected for $H^{\pm}$
from the CKM matrix $V$ and $L_2$ contains the flavor-changing effects for neutral
bosons as well as small additional flavor-changing terms for $H^{\pm}$.
The factor $\xi_{f_i}m_{f_i}$ and $\mu_{ij}^{f}$ arise primarily from the diagonal
and off-diagonal elements of $\Gamma^f_i$ respectively.

There are four major sources of CP violation\cite{WW,YLW}: (1) CKM matrix,(2) the phases
in factors $\xi_{f_i}$ which provide CP violation in charged-Higgs boson exchange;
(3) the phases in $\mu_{ij}^f$, this yields CP violation in FCNC.
(4) CP violation  in mixing 
matrix $O^H$. One of the most distinctive features of these sources is that
the factors $\xi_{f_i}$  can provide CP violation in charged Higgs boson
exchange in addition to and independent of CKM phase. 
As a consequence, in $\Delta S=1$ transitions its contribution
to $\epsilon'/\epsilon$ could be as large as $10^{-3}$ and become comparable with the experimental 
data\cite{NA31,KTEV}. Thus a measurement of 
$\epsilon'/\epsilon$ would not necessarily be due to CKM mechanism\cite{WW,YLW,YLW1}. 
This paper is organized as follows: In section 2, the analysis of the possible constraints of the 
Yukawa couplings from  
$F^0-\bar{F}^0$ mixings is presented. The various sources of CP violation and their influence 
on the determination of unitarity triangle are discussed in section 3, and section 4 contains 
our conclusions.
\def \beq{\begin{equation}}  \def \eeq{\end{equation}}
\def \beqa{\begin{eqnarray}} \def\eeqa{\end{eqnarray}}
\def \nonb{\nonumber}
\section{Constraints from $K^0-\bar{K}^0, B^0-\bar{B}^0$ and $D^0-\bar{D}^0$ mixings}

In the standard model, it is known that the neutral meson mixings arise
from the box diagram through two-W-boson exchange. The extremely small values
of the neutral $K$ and $B$ mass differences impose severe constraints on new physics
beyond the SM, especially on those with FCNC at tree level. In the S2HDM, additional
 contributions to the neutral meson mixings can arise from the box diagrams with 
charged-scalar exchanges and tree diagrams with neutral-scalar exchanges. 
The mass difference of $K_L-K_S$ is given by

\begin{equation}
\Delta m_{K} \simeq 2Re M_{12} \equiv 2Re (M_{12}^{WW} + M_{12}^{HH} +
M_{12}^{HW} + M_{12}^{H^{0}} + M'_{12} )
\end{equation}
where $M_{12}^{WW}$,  $M_{12}^{HH}$ and $M_{12}^{HW}$ are the contributions
from box diagrams through two $W^{\pm}$- boson, two charged-scalar $H^{\pm}$  and
one $W$- boson and one charged-scalar exchanges respectively.
$M_{12}^{H^{0}}$ is the one from the FCNC
through neutral scalar exchanges at tree level . $M'_{12}$ presents other possible contributions,
such as two-coupled penguin diagrams and nonperturbative effects.  They are resulted from the
corresponding effective Hamiltonian
\begin{eqnarray}
H_{eff}^{WW} & = & -\frac{G_F^{2}}{16\pi^{2}} m_{W}^{2} \sum_{i,j}^{c,t}
\eta_{ij}\lambda_{i} \lambda_{j} \sqrt{x_{i}x_{j}} B^{WW}(x_{i}, x_{j})
\bar{d}\gamma_{\mu}(1-\gamma_{5})s \bar{d}\gamma^{\mu}(1-\gamma_{5})s \\
H_{eff}^{HH} & = & -\frac{G_F^{2}}{16\pi^{2}} m_{W}^{2} \sum_{i,j}^{u,c,t}
\eta_{ij}^{HH} \lambda_{i} \lambda_{j} \frac{1}{4}\{ B_{V}^{HH}(y_{i}, y_{j})
[ \sqrt{x_{i}x_{j}}\sqrt{y_{i}y_{j}} |\xi_{i}|^{2}|\xi_{j}|^{2} \nonumber \\
& & \cdot \bar{d}\gamma_{\mu}(1-\gamma_{5})s \bar{d}\gamma^{\mu}
(1-\gamma_{5})s  + \sqrt{x_{s}x_{d}}\sqrt{y_{s}y_{d}}
\xi_{s}^{2}\xi_{d}^{\ast 2} \bar{d}\gamma_{\mu}(1+\gamma_{5})s
\bar{d}\gamma^{\mu}(1+\gamma_{5})s  \nonumber \\
& & + 2 \delta_{ij} \sqrt{x_{i}x_{j}} \sqrt{y_{s}y_{d}}
\xi_{s} \xi_{d}^{\ast} \xi_{i} \xi_{j}^{\ast} \bar{d}\gamma_{\mu}
(1+\gamma_{5})s \bar{d}\gamma^{\mu}(1-\gamma_{5})s ] \\
& & + B_{S}^{HH}(y_{i}, y_{j}) \sqrt{x_{i}y_{j}}[ x_{d} \xi_{d}^{\ast 2}
\xi_{i}^{\ast} \xi_{j}^{\ast} \bar{d}(1-\gamma_{5})s \bar{d}(1-\gamma_{5})s
\nonumber \\
 & & + x_{s}\xi_{s}^{2}\xi_{i}\xi_{j} \bar{d}(1+\gamma_{5})s
\bar{d}(1+\gamma_{5})s + 2 \sqrt{x_{s}x_{d}}\xi_{s} \xi_{d}^{\ast}
\xi_{i} \xi_{j}^{\ast} \bar{d}(1+\gamma_{5})s \bar{d}(1-\gamma_{5})s ] \}
\nonumber \\
H_{eff}^{HW} & = & -\frac{G}{16\pi^{2}} m_{W}^{2} \sum_{i,j}^{u,c,t}
\eta_{ij}^{HW}\lambda_{i} \lambda_{j} \{ 2 \sqrt{x_{i}x_{j}}\sqrt{y_{i}y_{j}}
\xi_{i}\xi_{j}^{\ast} B_{V}^{HW}(y_{i}, y_{j}, y_{w}) \nonumber \\
& & \cdot \bar{d}\gamma_{\mu}
(1-\gamma_{5})s \bar{d}\gamma^{\mu}(1-\gamma_{5})s  + (y_{i}+y_{j})
\sqrt{x_{d}x_{s}} \xi_{s}\xi_{d}^{\ast} [B_{T}^{HW}(y_{i}, y_{j},y_{w}) \\
& & \cdot \bar{d}\sigma_{\mu\nu}(1-\gamma_{5})s \bar{d}\sigma^{\mu\nu}
(1+\gamma_{5})s  + B_{S}^{HW}(y_{i},y_{j},y_{w}) \bar{d}(1-\gamma_{5})s
\bar{d}(1+\gamma_{5})s ] \} \nonumber
\end{eqnarray}
where the $B^{WW}$, $B_{V}^{HH}$, $B_{S}^{HH}$, $B_{V}^{HW}$, $B_{S}^{HW}$
and $B_{T}^{HW}$ arise from the loop integrals, and they are the functions of
$x_{i}=m_{i}^{2}/m_{W}^{2}$ and $y_{i}=m_{i}^{2}/m_{H}^{2}$ with $i=u,c,t,W$,
their explicit expressions are presented in the Appendix. Here $\eta_{ij}$,
$\eta_{ij}^{HH}$ and $\eta_{ij}^{HW}$  are the possible QCD corrections
and $\lambda_{i} = V_{is} V_{id}^{\ast}$.
Note that in obtaining above results
the external momentum of the d- and s-quark has been neglected. Except this
approximation which should be reliable as their current masses are small, we need to 
keep all the terms. This is because all the couplings $\lambda_{i}$ and $\xi_{i}$ are
complex in the model,  even if some terms are small, they can still play an
important role on CP violation since the observed CP-violating effect in
kaon decays is of order $10^{-3}$. The contributions of neutral Higgs bosons exchange 
at tree level can be evaluated by
\begin{eqnarray}
M_{12}^{H^{0}} & = & <P^0 | H_{eff}^{H^{0}} | \bar{P}^{0} > \nonumber \\
& = & \frac{G_F^{2}}{12\pi^{2}} f_{P^{0}}^{2} \tilde{B}_{P^{0}}m_{P^{0}}
(\sqrt{\frac{m_{f_{i}}}{m_{f_{j}}}})^{2}
(1+\frac{m_{f_{i}}}{m_{f_{j}}})^{-1} m_{f'_{j}}^{2}\sum_{k}
(\frac{2\sqrt{3} \pi v m_{P^{0}}} {m_{H_{k}^{0}}m_{f'_{j}}}
)^{2} (Y_{k,ij}^{f})^{2}
\end{eqnarray}
with

\[ (Y_{k,ij}^{f})^{2}  =  (Z_{k,ij}^{f})^{2} + \frac{1}{2}r_{P^{0}}
S_{k,ij}^{f} S_{k,ji}^{f \ast}, \qquad Z_{k,ij}^{f} =
-\frac{i}{2} (S_{k,ij}^{f} - S_{k,ji}^{f \ast}) \]

$S_{k,ij}$ is related to $\mu_{ij}^f$ through:
\beq
 S^f_{k,ij}=(O^H_{1k}+i \sigma_f O^H_{3k}) \frac{\mu^f_{ij}}{\sqrt{m_i m_j}} 
\eeq

where $\sigma_f=1$ for d-type quarks, $\sigma_f=-1$ for u-type quarks. 
The formula is expressed in a form which is convenient in comparison
with the one obtained from the box diagram in the SM, here
$\sqrt{m_{f_{i}}/m_{f_{j}}}$ with convention $i<j$ plays the role of
the CKM matrix element $V_{ij}$, and $m_{f'_{j}}$ is introduced to
correspond to the loop-quark mass of box diagram. Namely $f'_{j}$ and
$f_{j}$ are the two quarks in the same weak isospin doublet.
 Note that the result is actually independent of $m_{f'_{j}}$. 
Here $m_{f_{i}}$ are understood as the
current quark masses. In our following numerical estimations we will use
$m_{u}= 5.5$ MeV, $m_{d}= 9$ MeV, $m_{s}= 180$ MeV, $m_{c}= 1.4$ GeV  and
$m_{b}= 6$ GeV with being defined at a renormalization
scale of $1$ GeV. $f_{P^{0}}$ and $m_{P^{0}}$ are the leptonic decay
constant (with normalization $f_{\pi}=133$MeV) and the mass of the meson
$P^{0}$ respectively. $\tilde{B}_{P^{0}}$ and $\tilde{r}_{P^{0}}$ are bag
parameters defined via
\begin{eqnarray}
& & <P^0 |(\bar{f}_{i}(1\pm \gamma_{5})f_{j})^{2} | \bar{P}^{0} > =
- \frac{f_{P^{0}}m_{P^{0}}^{3}}{(m_{f_{i}}+m_{f_{j}})^{2}} \tilde{B}_{P^{0}} \\
& & 1 + \tilde{r}_{P^{0}} = - \frac{<P^0 |\bar{f}_{i}(1\pm \gamma_{5})f_{j}
\bar{f}_{i}(1\mp \gamma_{5})f_{j} | \bar{P}^{0} >}{<P^0 |\bar{f}_{i}(1\pm
\gamma_{5})f_{j} \bar{f}_{i}(1\pm \gamma_{5})f_{j} | \bar{P}^{0} >}
\end{eqnarray}
In the vacuum saturation and factorization approximation with the limit of
a large number of colors, we have  $\tilde{B}_{P^{0}}\rightarrow 1$ and
$\tilde{r}_{P^{0}}\rightarrow 0$, thus  $Y_{k,ij}^{f} = Z_{k,ij}^{f}$.

  It is known that $H_{eff}^{WW}$ contribution to $\Delta m_{K}$ is dominated
by the c-quark exchange and its value is still uncertain due to the large
uncertainties of the hadronic matrix element
\begin{equation}
 <K^0 |(\bar{d}\gamma_{\mu}(1-\gamma_{5})s)^{2} | \bar{K}^{0} > =
- \frac{8}{3} f_{K}^{2}m_{K}^{2} B_{K}
\end{equation}
where $B_{K}$ ranges from $1/3$ \cite{DGH1} (by the PCAC and $SU(3)$ symmetry),
$3/4$ \cite{BAG} (in the limit of a large number of colors) to $1$ \cite{GL}
(by the vacuum insertion approximation). The results from QCD sum rule and
Lattice calculations lie in this range. For small $B_{K}$, the
short-distance $H_{eff}^{WW}$ contribution to $\Delta m_{K}$ fails badly to
account for the measured mass difference. 
     
     In general, we obtain
\begin{eqnarray}
\Delta m_{K} & = & \frac{G_F^{2}}{6\pi^{2}} f_{K}^{2} B_{K}m_{K} m_{c}^{2}
\sin^{2} \theta \{ \eta_{cc} B^{WW}(x_{c}) + \frac{1}{4} \eta_{cc}^{HH} y_{c}
|\xi_{c}|^{4}  B_{V}^{HH}(y_{c}) \nonumber \\
& & + 2 \eta_{cc}^{HW} y_{c} |\xi_{c}|^{2}  B_{V}^{HW}(y_{c}, y_{w})
+\frac{\tilde{B}_{K}}{B_{K}} \sum_{k}
(\frac{2\sqrt{3} \pi v m_{K}} {m_{H_{k}^{0}}m_{c}}
)^{2} Re (Y_{k,12}^{d})^{2} \}
\end{eqnarray}
which is subject to the experimental constraint\cite{PDG}
\begin{equation}
\Delta m_{K} = 3.5 \times 10^{-6} eV \sim \sqrt{2}\frac{G_F^{2}}{6\pi^{2}} f_{K}^{2}
 m_{K} m_{c}^{2} \sin^{2} \theta  
\end{equation}
The effective Hamiltonian for  $B^{0}_{d}-\bar{B}^{0}_{d}$ Mixing is
calculated  with the aid of the box diagrams in full analogy to the
treatment of the $K^{0}-\bar{K}^{0}$ system. Its explicit expression can be
simply read off from the one for $K^{0}-\bar{K}^{0}$ system by a corresponding
replacement: $s\leftrightarrow b$. The "standard approximation" made there,
namely neglecting the external momenta of the quarks,  is also reliable
since dominant contributions come from the intermediate top quark.
With this analogy, the considerations and discussions on $K^{0}-\bar{K}^{0}$
mixing  can be applied to the $B^{0}_{d}-\bar{B}^{0}_{d}$
mixing for the contributions from box diagrams. As it is expected that
$|\Gamma_{12}|/2 \ll |M_{12}|$ in the B-system (which is different from
K-system), the mass difference for $B^{0}_{d}-\bar{B}^{0}_{d}$ system is
given by $\Delta m_{B} \simeq 2 |M_{12}|$.

 The general form for the mass difference in the $B^{0}_{d}-\bar{B}^{0}_{d}$
system  can be written
\begin{eqnarray}
\Delta m_{B} & \simeq & \frac{G_F^{2}}{6\pi^{2}} (f_{B}\sqrt{B_{B}\eta_{tt}})^{2}
m_{B} m_{t}^{2} |V_{td}|^{2} \frac{1}{\eta_{tt}} |\{ \eta_{tt} B^{WW}(x_{t})
+ \frac{1}{4} \eta_{tt}^{HH} y_{t} |\xi_{t}|^{4}  B_{V}^{HH}(y_{t}) \\
& & + 2 \eta_{tt}^{HW} y_{t} |\xi_{t}|^{2} B_{V}^{HW}(y_{t}, y_{w}) ]
+ \frac{\tilde{B}_{B}}{B_{B}} \sum_{k}
(\frac{2\sqrt{3} \pi v m_{B}} {m_{H_{k}^{0}}m_{t}}
)^{2} \frac{m_{d}}{m_{b}}\frac{1}{V_{td}^{2}}
 (Y_{k,13}^{d})^{2} \}| \nonumber
\end{eqnarray}
which is subject to the experimental constraint\cite{PDG}
\begin{equation}
\Delta m_{B} = (3.1 \pm 0.12)\times 10^{-4} eV \sim \frac{G_F^{2}}{12\pi^{2}}
(125MeV)  ^{2}  m_{B} (176GeV)^{2} (\sin\theta =0.22)^{6}
\end{equation}

    It is known that in the standard model the short-distance  contribution
to $\Delta m_{D}$ from the box diagram with W-boson exchange is of order
of magnitude  $\Delta m_{D}^{Box} \sim O(10^{-9})$ eV, here the external
momentum effects have to be considered and were found to suppress the
contribution by two orders of magnitude\cite{DK}. This is because of the
low mass of the intermediate state. It is not difficult to see that the
additional box diagram with charged-scalar exchange gives even smaller contribution
except $|\xi_{s}|$ is as large as $|\xi_{s}|\sim 2 m_{H^{+}}/m_{s}$ which
is unreliable large for the present bound $m_{H^{+}}> 54.5$ GeV\cite{PA}.  It has been
shown that dominant contribution to $\Delta m_{D}$ may come from the
long-distance effect since the intermediate states in the box diagram are
$d$- and $s$-quraks. The original estimations were found that
$\Delta m_{D}\sim 3 \times 10^{-5}$ eV \cite{WOLF2} and  $\Delta m_{D}\sim 1
\times 10^{-6}$ eV \cite{DGH2}.  An alternative calculation\cite{HG} using the heavy quark
effective theory showed that large cancellations among the intermediate states
may occur so that the long-distance standard model contribution to
$\Delta m_{D}$ is only larger by about one order of magnitude than the short-
distance contribution, which was also supported in a subsequent calculation
\cite{ORS}.

    With this in mind, we now consider the contribution to $\Delta m_{D}$
from the neutral scalar interaction in the S2HDM. It is easy to read off  from eq.(15)
\begin{eqnarray}
\Delta m_{D}^{H} & = & 2 |M_{12}^{H}|
=\frac{G_F^{2}}{6\pi^{2}} f_{D}^{2} \tilde{B}_{D}m_{D}
(\sqrt{\frac{m_{u}}{m_{c}}})^{2} m_{s}^{2}\sum_{k}
(\frac{2\sqrt{3} \pi v m_{D}} {m_{H_{k}^{0}}m_{s}}
)^{2} |Y_{k,12}^{u}|^{2} \nonumber \\
& = & 0.64 \times 10^{-4} (\frac{f_{D}\sqrt{\tilde{B}_{D}}}{210 MeV})^{2}
\sum_{k=1}^{3}(\frac{500 GeV}
{m_{H^{0}_{k}}})^{2} \frac{|Y_{k,12}^{u}|^{2}}{1}
\end{eqnarray}
With the above expected values in the second line for various parameters,
the predicted value for $\Delta m_{D}$ can be closed to the
current experimental limit $|\Delta m_{D}| < 1.58 \times 10^{-4}$ eV\cite{PDG},
this implies that a big $D^{0}-\bar{D}^{0}$ mixing which is larger than the
standard model prediction is not excluded. With this analysis, we come to the
conclusion that a positive signal of neutral
$D$ meson mixing from the future experiments at Fermilab, CESR at Cornell and
at a $\tau$-charm factory would be in favor of the S2HDM especially when the
exotic neutral scalars are not so heavy.

	We now proceed to study the constraints on the parameters
of the model. Since the parameters $\xi_{f_i}$ and $\mu^f_{ij}$ are in general all free
parameters, for simplicity we will consider the constraints in two  extreme
cases.

 Case 1: the mass difference is purely explained through the additional
box diagrams from two scalar-boson and one W-boson one scalar-boson. In this
case, the parameter $\xi_{f_i}$ is of particular importance. Both its 
amplitude and phase will play an important role in the neutral meson mass difference
and CP violation. It is quite different from the earlier analysis in type 1
and  type 2 2HDM \cite{Gilman} in which the three couplings $\xi_u, \xi_c,$ and $\xi_t$ are
equal, i.e., $\xi_u=\xi_c=\xi_t=\tan\beta$
. This is why the  constraint from $\epsilon$ is much stronger than the
one from $\Delta m_K$ in those models. In the S2HDM, one has
in general $\xi_u \neq \xi_c \neq \xi_t$, there are more degrees of freedom to 
fit $\epsilon$ and $\Delta m_K$, $\Delta m_B$ as well as $\Delta m_D$ .
   Since the main contribution to $\Delta m_K$ comes from c-quark though the
loop,  the upper bound of $\xi_c$ can be extracted from  $K^0-\bar{K}^0$ mixing.
The result is plotted in Fig.1. In the numerical calculations we take $f_K=161 $ MeV
and $B_K=0.75$. The rang of $m^+_H$ is from 100 to 500 GeV. Since the bound
of $|\xi_c|$ strongly depends on the SM prediction on $\Delta m_K$, three different
values of $\Lambda_{QCD}(\Lambda_{QCD}=0.21,0.31,0.41$) have been used, the corresponding ratios
to the experimental data $(\Delta m_K)_{exp}$ are 0.52,0.67 and 0.91\cite{herrlich}
  
 In the $B$ system, It is of interest to study its relative ratio to the SM, since large 
degree of uncertainty can be avoided from CKM matrix $|V_{td}|$ and hardonic
matrix elements. In Fig.2 we illustrate the  relation between $|\xi_t|$ and the charged
Higgs mass $m^+_H$ when the ratios of the HW and HH box diagram contributions to the
WW-box diagram contribution in the SM are 2:1, 1:1 and 0.5:1.  

  As it is shown in Fig.1 and Fig.2, in general $|\xi_c|$ is much larger than $|\xi_t|$.
Even when the HW and HH contributions to $\Delta m_B$ are twice as much as the SM one,
$|\xi_c|$ can still be larger than $|\xi_t|$ by an order of magnitude.

\newpage
  
\begin{figure}
\centerline{
\psfig{figure=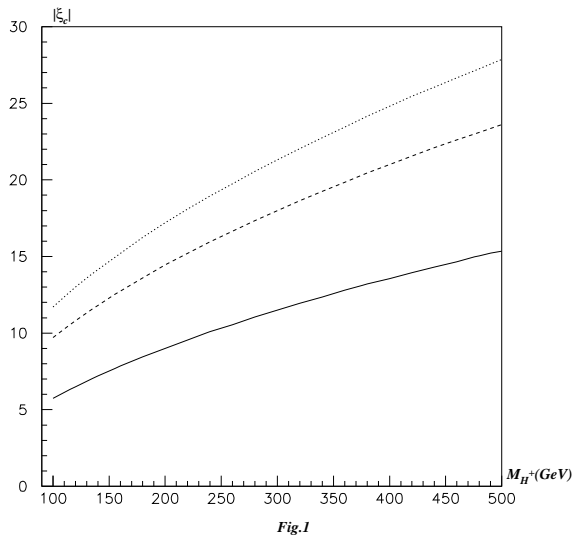, width=8 cm}
}
\caption{
The upper bound of $|\xi_c|$ with respect to the mass of the charged Higgs
scalar in case 1. The three curves are corresponding to the ratios 
$(\Delta m_K)_{SM}/(\Delta m_K)_{exp}$ = 0.52 (dotted), 0.67 (dashed) 
0.91 (solid) 
}
\end{figure}

\begin{figure}
\centerline{
\psfig{figure=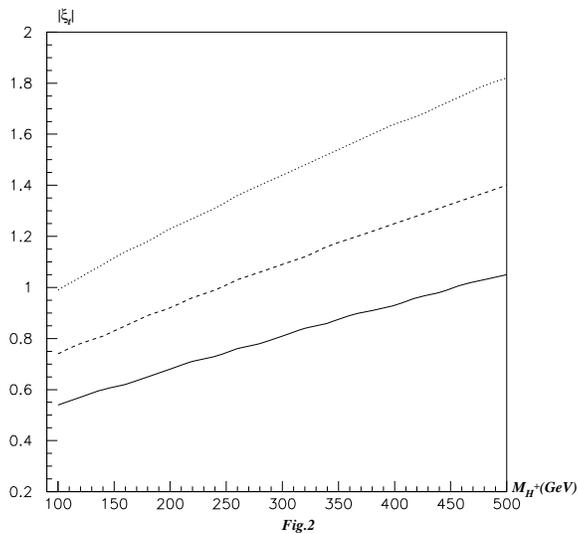, width=8 cm}
}
\caption{
The value of $|\xi_t|$ with respect to the mass of charged Higgs $m^+_H$.
The three curves are corresponding to the different ratios of HW and HH box diagram to the one
from the SM: 2:1 (dotted), 1:1 (dashed), 0.5:1 (solid).
}
\end{figure}

\newpage

 Case 2: the mass difference is fitted through neutral-scalar exchange at 
tree level. From Eq.(15) we know that the parameters arise in $M_{12}^0$ are 
$Y^f_{k,ij}$ rather then $\mu^f_{ij}$. If $S^f_{ij}$ is expected to be symmetric 
under the exchange $i \leftrightarrow j$ and $r_{P^0}=1$,  $Y^f_{k,ij}$ gets the
following simple form
\beq
Y^f_{k,i,j}=O^H_{1k} \frac{Im \mu^f_{ij}}{\sqrt{m_i m_j}}+
            \sigma_f O^H_{3k} \frac{Re \mu^f_{ij}}{\sqrt{m_i m_j}}
\eeq
 
Hence  both imaginary  and real part of  $\mu^f_{ij}$ are of  importance.
Futher more, the phases in $\mu^f_{ij}$ also provide a new source of CP violation 
as we have mentioned in the previous section.  To simplify the
discussions, we assume that one of the scalar-bosons, for example, the scalar $h$
 is much lighter than the other two scalars $H$ and $A$. The scalar bosons
 $H$ and $A$ are assumed to be heavier than 500 GeV.
 The upper bounds can be obtained from $K^0-\bar{K}^0, B^0-\bar{B}^0$ and
 $ D^0-\bar{D}^0$ mixings. The present consideration is more general than 
 the one in \cite{soni} where all the couplings $Y^f_{k,ij}$ are setted to be equal. As a consequence, 
 the constraints from different meson mixings provide different upper bounds upon different
 $Y^f_{k,ij}$. The results are shown in Fig.3. It is seen from the fig.3  that the upper
 bound of $Y^u_{k,12}$ is much higher than that from $K^0$ and $B^0$ system. This implies
 that a larger $ D^0-\bar{D}^0$ mixing than the standard model prediction
 is possible.

 \begin{figure}
\centerline{
\psfig{figure=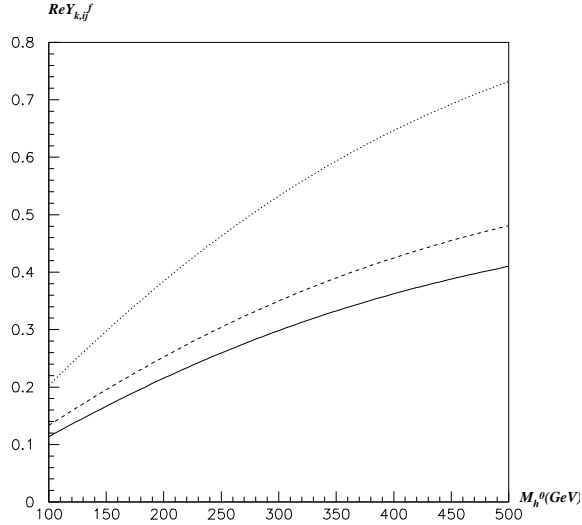, width=8 cm}
}
\caption{
The $m^0_h$ dependence of the upper bound of $ReY^f_{k,ij}$. 
$ReY^d_{1,12}$ from $K^0-\bar{K}^0$(solid),
$ReY^d_{1,13}$ from $B^0-\bar{B}^0$(dashed)
and $ReY^u_{1,12}$ from $D^0-\bar{D}^0$(dotted). The mass of the other scalar $m^0_A$
is fixed at $m^0_A=500$ GeV.  
}
\end{figure}

\section{ CP violation and unitarity triangle }
Besides the neutral meson mass difference, the indirect CP violation
parameter $\epsilon_K$ could also provide constraints on the values of $\xi_{f_i}$ and
$\mu^f_{ij}$. The standard definition of $\epsilon$ is
\begin{equation}
\epsilon = \frac{1}{\sqrt{2}}(\frac{Im M_{12}}{2 Re M_{12}} + \xi_{0})
e^{i\pi /4}
\end{equation}
where $\xi_{0}=Im A_{0}/ Re A_{0}$ with $|A_{0}| = (3.314\pm 0.004) \times
10^{-7}$ GeV the isospin-zero amplitude of $K\rightarrow \pi \pi $ decay.
Usually, the $\xi_{0}$ term is relatively small as it is proportional to
the small direct CP-violating parameter $\epsilon'$.

The first part of contribution to $\epsilon$ comes from the box diagram
through W-boson and charged-scalar exchanges
\begin{eqnarray}
Im M_{12}^{Box} & = & Im M_{12}^{WW} + Im M_{12}^{HH} + Im M_{12}^{HW}
\nonumber \\
 & = & \frac{G^{2}}{12\pi^{2}} f_{K}^{2} B_{K}m_{K} m_{i} m_{j}
\{ \sum_{i,j}^{c,t} Im (\lambda_{i} \lambda_{j}) Re B_{ij}(m_{i}, m_{j};
\xi_{i}, \xi_{j})  \nonumber \\
& & +  Re (\lambda_{i} \lambda_{j}) Im B_{ij}(m_{i}, m_{j}; \xi_{i}, \xi_{j})
\}
\end{eqnarray}
where $B_{ij}(m_{i}, m_{j}; \xi_{i}, \xi_{j})$ depend on the integral
functions of the box diagrams and their general form is given
in the Appendix. The imaginary part $Im B_{ij}(m_{i}, m_{j}; \xi_{i},
\xi_{j})$ arises from the complex couplings $\xi_{i}$.

 The second part is due to the flavor changing neutral scalar inteactions
at tree level
\begin{equation}
Im M_{12}^{H^{0}} = \frac{G^{2}}{12\pi^{2}} f_{K}^{2} \tilde{B}_{K}m_{K}
(\sqrt{\frac{m_{d}}{m_{s}}})^{2} m_{c}^{2}\sum_{k}
(\frac{2\sqrt{3}\pi v m_{K}} {m_{H_{k}^{0}}m_{c}})^{2} Im (Y_{k,12}^{d})^{2}
\end{equation}
This provides a contribution to $\epsilon$ in almost
any models which possess CP-violating flavor changing neutral scalar inteactions.

  In particular, the parameter $\epsilon$ could
receive large contributions from the long-distance dispersive effects
through the $\pi$, $\eta$ and $\eta'$ poles \cite{DPD}.
For a quantitative estimate of these effects, we follow the analyses in refs.
\cite{DPD,DH,HYC1,HYC2}
\begin{eqnarray}
(Im M'_{12})_{LD} & = & \frac{1}{4m_{K}} \sum_{i}^{\pi,\eta, \eta'} \frac{Im
(<K^{0}|L_{eff}|i><i|L_{eff}|\bar{K}^{0}>)}{m_{K}^{2} - m_{\pi}^{2}}
\nonumber  \\
 & = & \frac{1}{4m_{K}} \frac{2\kappa}{m_{K}^{2} - m_{\pi}^{2}}
<K^{0}|L_{-}|\pi^{0}><\pi^{0}|L_{+}|\bar{K}^{0}>) \\
 & = & \frac{G^{2}}{12\pi^{2}} f_{K}^{2} B'_{K}m_{K} (\frac{m_{K}}
{m_{s}})^{2} \sin\theta m_{s}^{2} (\sqrt{\frac{\pi \alpha_{s}}{2}}
\frac{3\kappa A_{K\pi}}{4m_{s}(m_{K}^{2} - m_{\pi}^{2})}) \nonumber \\
& & \cdot \sum_{i} [Im \lambda_{i} Re P_{i}^{H}(m_{i}, \xi_{i}) + Re \lambda_{i}
Im P_{i}^{H}(m_{i}, \xi_{i})]
\end{eqnarray}
where $\kappa$ is found to be $\kappa \simeq 0.15$ when considering the
$SU(3)-$ breaking  effects in the $K-\eta_{8}$ transition, and nonet-symmtry-
breaking in $K-\eta_{o}$ as well as $\eta-\eta'$ mixing. We shall not repeat
these analyses, and the reader who is interested in it is refered to the
paper \cite{HYC2} and references therein. $L_{-}$ and $L_{+}$ are CP-odd and
CP-even lagrangians respectively (with convention $L_{eff}= L_{+} + i L_{-}$).
The $L_{-}$ is induced from the gluon-penguin diagram with charged-scalar
\begin{equation}
L_{-} = f_{s} \bar{d}\sigma_{\mu\nu}(1+\gamma_{5})\lambda^{a} s G_{\mu\nu}^{a}
 - f_{d} \bar{d}\sigma_{\mu\nu}(1-\gamma_{5})\lambda^{a} s G_{\mu\nu}^{a}
\end{equation}
with
\begin{equation}
f_{q} = \frac{G}{\sqrt{2}} \frac{g_{s}}{32\pi^{2}} m_{q} \sum_{i}
Im(\xi_{q}\xi_{i} \lambda_{i}) y_{i} P_{T}^{H}(y_{i})
\end{equation}
where $P_{T}^{H}(y_{i})$ is the integral function and presented in the
Appendix. From $f_{s}$ and $f_{d}$ it is not difficult to read off the
$Re P_{i}^{H}(m_{i}, \xi_{i})$ and $Im P_{i}^{H}(m_{i}, \xi_{i})$ (see Appendix).
In obtaining the last expression of the above equation, we have used the
result for the hadronic matrix element $<K^{0}|L_{-}|\pi^{0}> = (f_{s} - f_{d}) A_{K\pi}$ 
where $A_{K\pi}$ has been computed in the MIT bag model and was found \cite{DHH} to be $A_{K\pi}
= 0.4 GeV^{3}$ for $\alpha_{s}=1$, and the convention
$<\pi^{0}|L_{+}|\bar{K}^{0}>=\frac{1}{2} G f_{K}^{2} B'_{K}m_{K}^{2}
(2m_{K}/m_{s})^{2} \sin\theta $, where $B'_{K}$ is introduced to
fit the experimental value $<\pi^{0}|L_{+}|\bar{K}^{0}>= 2.58 \times 10^{-7}
GeV^{2}$ and is found to be $B'_{K}=1.08$. We then obtain
$\sqrt{\pi \alpha_{s}} 3\kappa A_{K\pi}/[4\sqrt{2}m_{s}(m_{K}^{2} -
m_{\pi}^{2})] \simeq 1.4$.

Neglecting the $t-$ and $u-$quark contributions and also the terms
proportional to $m_{d}$ in comparison with the terms proportional to $m_{s}$,
the total contributions to the CP-violating parameter $\epsilon$ can be
simply calaculated from the following formula
\begin{eqnarray}
|\epsilon| & = & 3.2\times 10^{-3} B_{K} (\frac{|V_{cb}|}{0.04})^{2}
\frac{2|V_{ub}|}{|V_{cb}||V_{us}|}  \sin\delta_{KM} \{ -\frac{1}{4}
[\eta_{cc} B^{WW}(x_{c}) \nonumber \\
& & + \frac{1}{4} \eta_{cc}^{HH} y_{c}
|\xi_{c}|^{4}  B_{V}^{HH}(y_{c}) + 2 \eta_{cc}^{HW} y_{c} |\xi_{c}|^{2}
B_{V}^{HW}(y_{c}, y_{w})] \nonumber \\
& & + (\frac{|V_{cb}| m_{t}}{2m_{c}})^{2}(1-\frac{|V_{ub}|}{|V_{cb}||V_{us}|}
\cos\delta_{KM}) [\eta_{tt} B^{WW}(x_{t}) \nonumber \\
& & + \frac{1}{4}
\eta_{tt}^{HH} y_{t} |\xi_{t}|^{4}  B_{V}^{HH}(y_{t}) + 2 \eta_{tt}^{HW}
y_{t} |\xi_{t}|^{2}  B_{V}^{HW}(y_{t}, y_{w})] \nonumber \\
& & + \frac{m_{t}}{4m_{c}}[\eta_{ct} B^{WW}(x_{c}, x_{t}) + \frac{1}{2}
\eta_{ct}^{HH} \sqrt{y_{c}y_{t}}
|\xi_{c}|^{2} |\xi_{t}|^{2} B_{V}^{HH}(y_{c}, y_{t}) \nonumber \\
& & + 4 \eta_{ct}^{HW} \sqrt{y_{c}y_{t}}Re(\xi_{c}\xi_{t})
B_{V}^{HW}(y_{c}, y_{t}, y_{w}) ]\}    \\
& &  + 2.27\times 10^{-3} \frac{Im (\tilde{Y}_{k,12}^{d})^{2}}{6.4\times
10^{-3}}\tilde{B}_{K}\sum_{k}(\frac{10^{3} GeV} {m_{H_{k}^{0}}})^{2}  \nonumber \\
& & + 2.27\times 10^{-3} Im(\xi_{c}^{\ast}\xi_{s}^{\ast})^{2} \frac{6.8 GeV^{2}}
{m_{H^{+}}^{2}}\tilde{B}_{K} (ln\frac{m_{H^{+}}^{2}}{m_{c}^{2}} -2) \nonumber
\\
& & + 2.27 \times 10^{-3} Im(\xi_{c}\xi_{s}) \frac{37 GeV^{2}}{m_{H^{+}}^{2}}
B'_{K} (ln\frac{m_{H^{+}}^{2}}{m_{c}^{2}} -\frac{3}{2})
+ \frac{\xi_{o}}{\sqrt{2}} \nonumber 
\end{eqnarray}
where we have used the experimental constraint on $2 Re M_{12}=
\Delta m_{K}^{exp.}$ 

  It is analogous to the section 2, we consider the contributions to $\epsilon$ 
in two different cases. The first one, CP violation is governed by 
the induced KM mechanism, i.e., the first term of the above equation becomes 
dominant. In this case, new contributions come from the box diagrams of
two charged-scalar, and one W-boson and one charged scalar exchanges. Since the
expression contains $Re(\xi_c\xi_t)$, the relative phase $\theta$ between $\xi_c$ and
$\xi_t$, i.e., $Re(\xi_c\xi_t)=|\xi_c||\xi_t|\cos\theta$,  
may play an important role. It is of interest to illustrate how such effects
can influence  the determination of the unitarity triangle. In Fig.4, the constraint
to the vertex of the unitarity triangle is given from $|V_{ub}|,\Delta m_B$ and $\epsilon$. 
Here the new physics effect can change the  value of $|V_{td}|$ and the
shape of the bounds from $\epsilon$. It is different from the ref.\cite{BCK} in which 
$|V_{td}|$ was fixed and taken the value  $|V_{td}|=0.0084$.  In our numerical calculations, 
we take $|\xi_c|=9.8$ and $|\xi_t|=0.54$, the mass of charged Higgs is fixed at $M^+_H=200$ GeV. 
The relative phase between them is taken to be $\pi/3$ and 
$2\pi/3$ as two examples. The other input parameters are: $B_K=0.75\pm0.15, |V_{ub}|/
|V_{cb}|=0.08\pm0.02$ and $|V_{cb}|=0.04$.
  For a comparison, we carry out a similar calculation
in the SM with the same parameters, the result is plotted in Fig.5. It is found
that the shape of the triangle can be largely changed when taking different values of the 
relative phase between $\xi_c$ and $\xi_t$. The angle $\beta$ of the triangle
may be extremely small when $\cos\theta$ is close to 1.  

    The second case, $\epsilon_K$ receives contributions from
both the charged-scalar and the neutral-scalar exchanges. 
This case is more important than the one in which 
only the neutral-scalar exchange is dominant. This is because the relative phase between the 
the two contributions can largely affect the determination of
$|V_{td}|$. To illustrate such phase effect, we choose the ratio between the charged- and 
neutral-scalar contributions to $\Delta m_B$ to be 2:1, and
take four typical values for the relative phase between them as $0, \pi/3, 2\pi/3$ and $\pi$. As it
was pointed out by Soares and Wolfenstein\cite{soares} that if such phase
emerges, then the unitarity angle extracted from $B \rightarrow J/\psi K_S$ will
be the total phase $\phi_M$ rather than $\beta$. Here $\phi_M$ is defined by
\[
 M_{12}^{total}=|M_{12}^{SM}+M_{12}^{NEW}|\exp^{2i\phi_M}
 \]
 with the index 'SM' and 'NEW' denoting the contributions from the SM
 and new physics. Recently, the preliminary measurement of $B\rightarrow J\psi K_{S}$ 
was reported by CDF Collaboration\cite{cdf} with $a_{\psi K_{S}}=0.79^{+0.41}_{-0.44}$. However, 
the resulting constraints on new physics are not very strong\cite{nir} due to the large errors. 
A more precise measurement is expected at B factories. 
 
  In Fig.6,  we plot the value of $V_{td}$
extracted from $\Delta m_B$ in the $\rho-\eta$ plane without considering
the uncertainty of $B_K( B_K=0.75)$. The four curves are corresponding to the above
four cases. The figure shows that the additional phase from $Y^f_{k,ij}$ can
strongly change the value of $V_{td}$. Its modulus varies in the interval between
0.7 and 1.2 in this situation.

 As an example, the influence on the determination of the unitarity
triangle in the case 3 (see fig. 6) is plotted in Fig.7. 
The bounds from $\epsilon$ become lower than that from the SM due to the relative phase effect.
As the three bounds from $|V_{ub}|, \Delta m_B$ and $\epsilon$ still have area in common, the
triangle remains to be closed. However, as we have mentioned above, when the angle
$\beta$ is extracted from $B \rightarrow J/\psi K_S$, its value will be the 
total phase $\phi_M$ which may be  much larger. As a consequence, it could make 
the unitarity triangle to be `open. This possibility has been shown to happen in 
the case 3, where $\tan\phi_M$ is three times as large as $\tan\beta$.

\begin{figure}
\centerline{
\psfig{figure=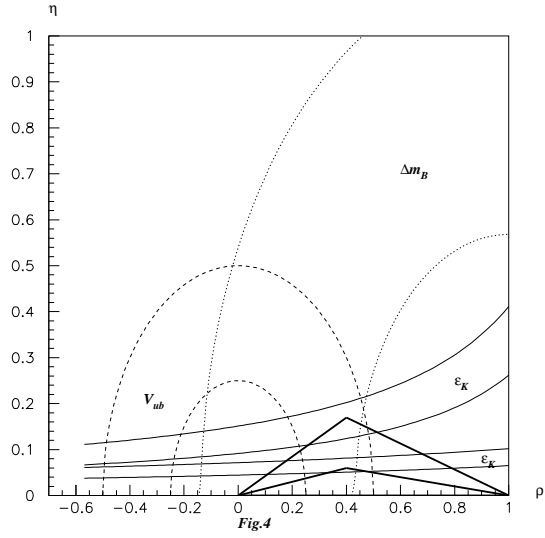, width=8 cm}
}
\caption{The constraints on the unitarity triangle in $\eta-\rho$ plane,
 the two different triangles
correspond to $\theta=\pi/3$ (case a) and $\theta=2\pi/3$ (case b). 
Where $\theta$ is the relative phase between 
$\xi_{c}$ and $\xi_{t}$. Other parameters are
 $B_K=0.75\pm0.15, |V_{ub}|/|V_{cb}|=0.08\pm0.02$ and $|V_{cb}|=0.04$. The allowed region between 
the two dashed curves are from $|V_{cb}|$, the allowed region between two dotted curves 
are from $\Delta m_{B}$ and the allowed region between two solid curves 
are from $|\epsilon_{K}|$ (the two curves below are for the case a for $\theta = \pi/3$ and the
two curves above are for the case b for $\theta = 2\pi/3$.)}
\end{figure}

\begin{figure}
\centerline{
\psfig{figure=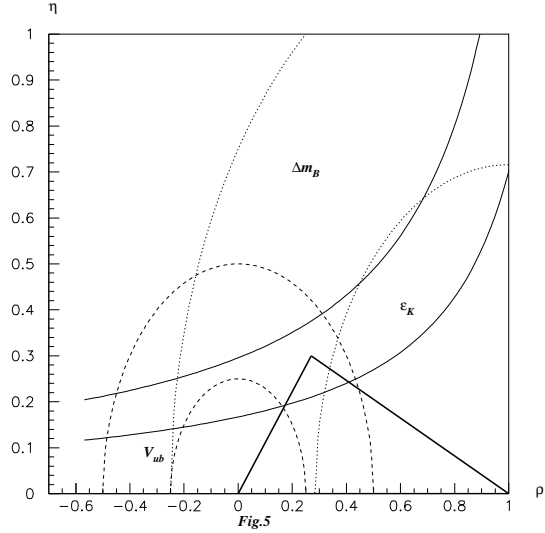, width=8 cm}
}
\caption{The constraints on the unitarity triangle from SM. The parameters $B_K,|V_{ub}|,
|V_{cb}| $ are as the same as in Fig.4 }
\end{figure}

\begin{figure}
\centerline{
\psfig{figure=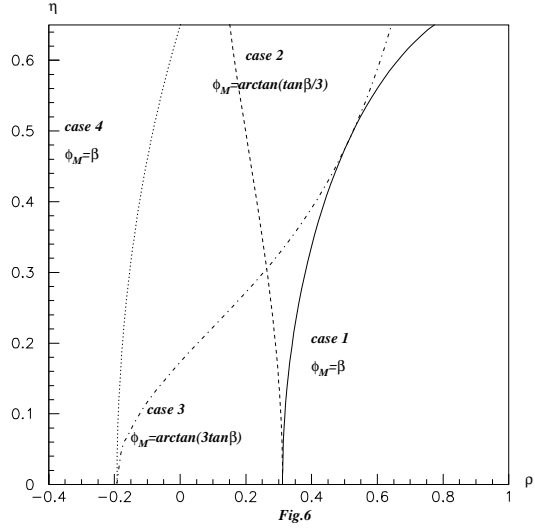, width=8 cm}
}
\caption{Constraints on $V_{td} $ from $\Delta m_B$. The relative phase between
charged- and neutral-scalar exchange is taken to be 0(case 1, solid),$\pi/3$(
case 2, dashed),$2\pi/3$(case 3, dash-dotted) and $\pi$(case 4, dotted)  }
\end{figure}

\begin{figure}
\centerline{
\psfig{figure=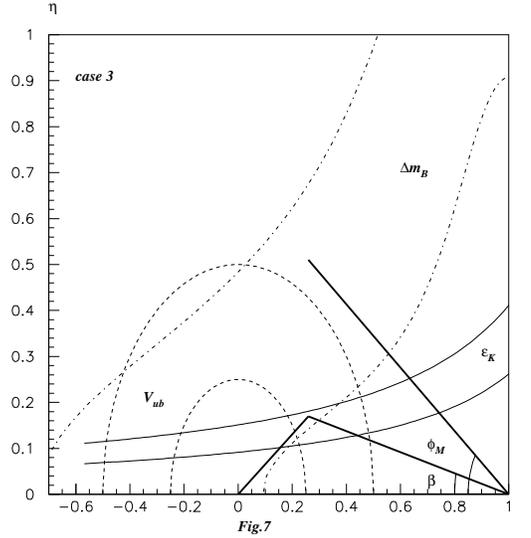, width=8 cm}
}
\caption{The constraints on the unitarity triangle in case 3, 
where $\tan\phi_M$ is three times as large as $\tan\beta$   }
\end{figure}

\section{CONCLUSIONS}
  
In conclusion, we have studied one of the simplest extentions of the standard model 
with an extra Higgs doublet, which we have simply called as an S2HDM, some 
constraints on the parameters in the S2HDM have been obtained from $F^0-\bar{F}^0$ mixing 
processes. It has been shown that in general $\xi_u \neq \xi_c \neq \xi_t$ and 
$|\xi_c| \gg |\xi_t| $. A much larger $D^0-\bar{D}^0$ mixing than the SM prediction is possible.
Various sources of CP violation have been discussed. 
Their influences on the determination of the unitarity triangle have also been studied in detail.
We found that angle $\beta$ of unitarity triangle could be largely suppressed 
due to the new contribution from Higgs box diagrams. The phase from neutral Higgs 
exchange could strongly affect the extraction of $\beta$ from $B\rightarrow J/\psi K_S$. 
In some cases, such an effect could be so large that the unitarity triangle cannot 
remain to be closed. In particular, it may even result in the angle $\beta$ determined from fitting 
the quantities $|V_{ub}|$, $\Delta m_{K}$ and $\epsilon$ being different from the one extracted 
from the decay process $B\rightarrow J/\psi K_S$.  If it is so, a clear signal of new physics is indicated. 

{\bf Acknowledgments:} This work was supported in part by the NSF of China under the grant No. 19625514.

\begin{center} {\bf Appendix} \end{center}

  In this appendix, we present some functions and quantities appearing in the
text.

A. \   The integral functions from the box diagrams with W boson and charged
scalar exchanges

\begin{eqnarray*}
   B^{WW}(x, x') & = & \sqrt{xx'}\{ \frac{1}{4} + \frac{3}{2}\frac{1}{1-x} -
\frac{3}{4}\frac{1}{(1-x)^{2}} \frac{\ln x}{x-x'} \\
& & + (x \leftrightarrow x') - \frac{3}{4} \frac{1}{(1-x)(1-x')} \} \\
   B^{WW}(x) & = & \frac{1}{4} + \frac{9}{4}\frac{1}{1-x} -
\frac{3}{2}\frac{1}{(1-x)^{2}} +\frac{3}{2}\frac{x^{2}}{(1-x)^{3}} \ln x \\
   B^{HH}_{V}(y, y') & = & 16\pi^{2} m_{H}^{2} I_{4} \\
 & = & \frac{y^{2}}{(y-y')(1-y)^{2}} \ln y + (y \leftrightarrow y')
+ \frac{1}{(1-y)(1-y')}  \\
 B^{HH}_{V}(y) & = & 16\pi^{2} m_{H}^{2} I_{1} \\
 & = & \frac{1+y}{(1-y)^{2}} +  \frac{2y}{(1-y)^{3}} \ln y  \\
   B^{HW}_{V}(y, y', y_{W}) & = & 16\pi^{2} m_{H}^{2} (\frac{1}{4}I_{6}
+ m_{W}^{2} I_{5}) =  \frac{(y_{W}-1/4)\ln y_{W}}{(1-y)(1-y')(1-y_{W})} \\
& & - \frac{y(y_{W}-y/4)}{(y-y')(1-y)(y-y_{W})} \ln \frac{y_{W}}{y} + (y
\leftrightarrow y')  \\
   B^{HW}_{V}(y, y_{W}) & = & 16\pi^{2} m_{H}^{2} (\frac{1}{4}I_{3}
+ m_{W}^{2} I_{2})  =  \frac{y_{W}-1/4 y}{(1-y)(y-y_{W})} \\
& & + \frac{(y_{W}-1/4)}{(1-y)^{2}(1-y_{W})} \ln y
+ \frac{3}{4}\frac{y_{W}^{2}}{(y_{W}-y)^{2}(1-y_{W})} \ln \frac{y_{W}}{y}
  \\
  B^{HH}_{S}(y, y') & = & -\frac{y\ln y}{(y-y')(1-y)^{2}} \ln y -
(y \leftrightarrow y') - \frac{1}{(1-y)(1-y')}  \\
 B^{HH}_{S}(y) & = & -\frac{1}{(1-y)^{2}}[\frac{1+y}{1-y}  \ln y + 2 ] \\
   B^{HW}_{S}(y, y', y_{W}) & = & 16\pi^{2} m_{H}^{2} I_{6} =
- \frac{\ln y_{W}}{(1-y)(1-y')(1-y_{W})} \\
& & +  \frac{y^{2}}{(y-y')(1-y)(y-y_{W})} \ln \frac{y_{W}}{y} + (y
\leftrightarrow y')   \\
 B^{HW}_{S}(y, y_{W}) & = & 16\pi^{2} m_{H}^{2} I_{3}  =
-\frac{y}{(1-y)(y-y_{W})} \\
& &  -\frac{\ln y}{(1-y)^{2}(1-y_{W})}
-\frac{y_{W}^{2}}{(y_{W}-y)^{2}(1-y_{W})} \ln \frac{y_{W}}{y}
  \\
   B^{HW}_{T}(y, y', y_{W}) & = & 16\pi^{2} m_{H}^{2} m_{W}^{2} I_{5} =
 y_{W} \{ \frac{\ln y_{W}}{(1-y)(1-y')(1-y_{W})}  \\
& & + \frac{y\ln y/y_{W}}{(y-y')(1-y)(y-y_{W})}  + (y
\leftrightarrow y')  \\
   B^{HW}_{T}(y, y_{W}) & = & 16\pi^{2} m_{H}^{2} m_{W}^{2} I_{2} =
y_{W} \{ \frac{1}{(1-y)(y-y_{W})} \\
& & + \frac{\ln y}{(1-y)^{2}(1-y_{W})}
- \frac{y_{W} \ln y/y_{W}}{(y_{W}-y)^{2}(1-y_{W})}
\end{eqnarray*}
where the functions $I_{i}$ ($i=1, \cdots , 6$) are the euclidean integrals\cite{ASW}:

\begin{eqnarray*}
I_1(m)& = &\frac{1}{16\pi^2}\left[ \frac{M_H^2+m^2}{M_H^2-m^2)^2} 
       +\frac{2m^2M_H^2}{M_H^2-m^2)^3} \ln\left(\frac{m^2}{M^2}\right)\right]\\
I_2(m)& = &\frac{-1}{16\pi^2M_W^2M_H^2} \left[ \frac{M_W^2\ln(M_H^2/m^2)}{
          (M_H^2-M_W^2)} + \frac{M_W^2\ln(M_W^2/m^2)}{(M_W^2-M_H^2)} +1 \right]\\
I_3(m)& = &\frac{1}{16\pi^2}\left[\frac{\ln(M_H^2/M_W^2)}{(M_H^2-M_W^2} \right]\\
I_4(m_i,m_j)& = & \frac{1}{16\pi^2}\left[
               \frac{m_i^4\ln(m_i^2/M_H)^2}{(m_i^2-m_j^2)(M_H^2-m_i^2)^2}
              +\frac{m_j^4\ln(m_j^2/M_H)^2}{(m_j^2-m_i^2)(M_H^2-m_j^2)^2}
              + \frac{M_H^2}{(M_H^2-m_i^2)(M_H^2-m_j^2)} \right] \\
I_5(m_i,m_j)& = & \frac{1}{16\pi^2}\left[
            \frac{M_H^2\ln(M_W^2/M_H^2)}{(M_H^2-M_W^2)(M_H^2-m_i^2)(M_H^2-m_j^2)}
           +\frac{m_i^2\ln(M_W^2/m_i^2)}{(m_i^2-M_W^2)(m_i^2-M_H^2)(m_i^2-m_j^2)}\right. \\
&  &\left. +\frac{m_j^2\ln(M_W^2/m_j^2)}{(m_j^2-M_W^2)(m_j^2-M_H^2)(m_j^2-m_i^2)}
            \right] \\
I_6(m_i,m_j)& = & \frac{1}{16\pi^2}\left[
            \frac{M_H^4\ln(M_H^2/M_W^2)}{(M_H^2-M_W^2)(M_H^2-m_i^2)(M_H^2-m_j^2)}
           +\frac{m_i^4\ln(m_i^2/M_W^2)}{(m_i^2-M_W^2)(m_i^2-m_H^2)(m_i^2-m_j^2)}\right. \\
 & & \left. +\frac{m_j^4\ln(m_j^2/M_W^2)}{(m_j^2-M_W^2)(m_j^2-m_H^2)(m_j^2-m_i^2)}
           \right] \\
\end{eqnarray*}       

B. \   The quantities used for calculating the CP-violating parameter
$\epsilon$.

\begin{eqnarray*}
B_{ij}(m_{i}, m_{j}) & = & \sqrt{y_{i}y_{j}}\{ \frac{1}{4}
 |\xi_{i}|^{2}|\xi_{j}|^{2} B_{V}^{HH}(y_{i}, y_{j})
+ 2  Re(\xi_{i}\xi_{j}^{\ast}) B_{V}^{HW}(y_{i}, y_{j}, y_{W}) \}
\\
& & + \eta_{ij} B^{WW}(x_{i}, x_{j}) + \frac{m_{s}^{2}}{4m_{i}m_{j}}
\{ \frac{\tilde{B}_{K}}{B_{K}}
(\frac{m_{K}}{(m_{d}+m_{s})})^{2} \sqrt{y_{i}y_{j}} B_{S}^{HH}(y_{i}, y_{j})
\xi_{ij}^{2}  \\
& & + 2 \frac{m_{d}}{m_{s}} \sqrt{y_{i}y_{j}} B_{V}^{HH}(y_{i}, y_{j})
[\xi_{s}\xi_{d}^{\ast}\xi_{i}\xi_{j}^{\ast} + \frac{1}{2} \sqrt{\frac{m_{d}
m_{s}}{m_{i}m_{j}}} (\xi_{s}\xi_{d}^{\ast})^{2}] \}
\end{eqnarray*}
with
\[ \xi_{ij}^{2} = (\xi_{i}\xi_{s} - \frac{m_{d}}{m_{s}}\xi_{i}^{\ast}
\xi_{d}^{\ast})(\xi_{j}\xi_{s} - \frac{m_{d}}{m_{s}}\xi_{j}^{\ast}
\xi_{d}^{\ast}) - 2 \tilde{r}_{K} \frac{m_{d}}{m_{s}} \xi_{i}\xi_{s}
(\xi_{j}\xi_{d})^{\ast} \]

The involved functions arise from the box graphs. From gluonic penguin graph
with
charged scalar exchange, we have

\[ P^{H}_{T}(y_{i}) = \frac{1}{2(1-y_{i})} + \frac{1}{(1-y_{i})^{2}}
+ \frac{1}{(1-y_{i})^{3}} \ln y_{i} \]

and

\[ P^{H}_{i}(m_{i}, \xi_{i}) = (\xi_{s} \xi_{i} - \frac{m_{d}}{m_{s}} \xi_{d}
\xi_{i}) y_{i} P^{H}_{T}(y_{i}) \]

\end{document}